# 1D Convolutional Neural Network Models for Sleep Arousal Detection

Morteza Zabihi[1], Ali Bahrami Rad[2], Serkan Kiranyaz[3], Simo Särkkä[2], Moncef Gabbouj[1]

[1] Department of Computing Sciences, Tampere University, Tampere, Finland
[2] Department of Electrical Engineering and Automation, Aalto University, Espoo, Finland
[3] Electrical Engineering Department, Qatar University, Doha, Qatar

E-mail: Morteza.zabihi@tuni.fi

**Abstract**

Sleep arousals transition the depth of sleep to a more superficial stage. The occurrence of such events is often considered as a protective mechanism to alert the body of harmful stimuli. Thus, accurate sleep arousal detection can lead to an enhanced understanding of the underlying causes and influencing the assessment of sleep quality. Previous studies and guidelines have suggested that sleep arousals are linked mainly to abrupt frequency shifts in EEG signals, but the proposed rules are shown to be insufficient for a comprehensive characterization of arousals. This study investigates the application of five recent convolutional neural networks (CNNs) for sleep arousal detection and performs comparative evaluations to determine the best model for this task. The investigated state-of-the-art CNN models have originally been designed for image or speech processing. A detailed set of evaluations is performed on the benchmark dataset provided by PhysioNet/Computing in Cardiology Challenge 2018, and the results show that the best 1D CNN model has achieved an average of 0.31 and 0.84 for the area under the precision-recall and area under the ROC curves, respectively.

## 1. Introduction

Sleep arousal is typically defined as a sleep disruption process which results in fragmented sleep (American sleep disorders association, 1992). Despite the substantial evidence of sleep arousal occurrences, various definitions and criteria are proposed for arousal detection (Halász et al 2004). The American Academy of Sleep Medicine (AASM) introduced a guideline as an attempt to unify arousal definitions (Rosenberg and Van Hout 2014). They defined arousal as an abrupt shift of EEG frequency, which can happen in theta, alpha, or higher frequency bands. Besides, AASM mentions that arousal can be associated with an increase in electromyographic or heart activity (Rosenberg and Van Hout 2014). This sleep scoring manual considers an arousal disruption to last at least three seconds. Moreover, the use of other measurements such as respiratory events is suggested for improvement in arousal scoring. Although the proposed AASM scoring is prevalent as a standard approach, it is strongly criticized by several research groups because of insufficient evidence for most of the proposed rules (Parrino et al 2009), and for not considering the time-varying nature of sleep (Schulz 2008).

Due to such issues and critics, data-driven models such as deep convolutional neural networks (CNNs) have recently become more commonly used in this field. One of the main advantages of end-to-end deep learning models compared to other machine learning algorithms is their minimal reliance on expert knowledge. This can be considered as a practical solution for the assessment of sleep arousals, specifically in this case, where there is a lack of comprehensive criteria for sleep arousal detection (Grigg-Damberger 2012). In an endeavor to motivate researchers to contribute in this domain, a sleep disorder dataset including 1985 subjects is provided by the PhysioNet Computing in Cardiology (CinC) Challenge 2018 (PhysioNet 2018). In this paper, we narrow our focus on this dataset, as it is the most recent and largest publicly available dataset for sleep arousal. The goal of the challenge was to use 13 polysomnography (PSG) signals to detect non-apnea arousals during sleep (i.e., discriminate between arousal and non-arousal events). Of all the 22 final entries, the majority of the teams used deep learning approaches as a fundamental element of their proposed algorithms. The top-scoring algorithms in this challenge can be categorized into two main groups: 1) end-to-end deep learning algorithms, and 2) deep learning algorithms combined with handcrafted features.

The proposed works by Howe-Patterson et al (2018) and He et al (2018) fall into the first group. They used 12 PSG channels and a network of multiple convolutional layers followed by a long-short-term memory (LSTM). The method proposed by Howe-Patterson et al (2018) achieved an impressive area under the precision-recall curve (AUPRC) of 0.54. One of the critical features of their proposed architecture is the use of dilated convolutions, which increase the receptive view of the network without the resolution loss. In several works, such as WaveNet (Oord et al 2016) and ByteNet (Kalchbrenner et al 2017), it has been shown that dilated convolution is effective in speech synthesis and machine translation. Furthermore, the CNN blocks introduced in Howe-Patterson et al (2018) and He et al (2018) take advantage of shortcut connections similar to the residual block proposed in ResNet (He et al 2016). In the last few years, several studies (e.g., Huang et al 2017 and Xie et al 2017) reported boosted performance using different variants of the ResNet architecture.

Among the top ten performers, Patane et al (2018) and Miller et al (2018) proposed end-to-end learning approaches using CNN architectures. Patane et al (2018) achieved AUPRC of 0.42 using an ensemble of CNNs, where each CNN is responsible for learning features from a specific channel. They used six layers of one-dimensional CNN for each channel and then fused them through a Siamese architecture (Bromley et al 1993). The model presented by Miller et al (2018) achieved AUPRC of 0.36 using a convolutional-deconvolutional network architecture. Often, encoder-decoder networks are used for segmentation problems such as UNet (Ronneberger et al 2015). Their proposed model utilizes shortcut connections to connect the input of each deconvolutional layer with the output of the corresponding convolutional layer. Such bypass connections are effective in training deeper networks by solving the gradient vanishing problem (He et al 2016).

Interestingly, the works proposed by Már-Þráinsson et al (2018) and Varga et al (2018) show that handcrafted features based on expert knowledge can perform in line with some of the top-scoring end-to-end deep learning algorithms. These works benefited from the combination of knowledge-based features and deep learning methods. In Már-Þráinsson et al (2018), handcrafted features were extracted from time, frequency, and time-frequency domains over a 10-second sliding window with 50% overlap. They used a three-layer model consisting of a bidirectional LSTM with 50 hidden units, a fully connected layer with 50 neurons, and a softmax layer as the output layer. AUPRC of 0.45 and 0.42 are achieved by Már-Þráinsson et al (2018) and Varga et al (2018), revealing the efficiency of their proposed methods.

In summary, due to the complexity of the problem and the amount of data in this challenge, most entries used deep learning approaches. In addition, employing most of the PSG measurements (i.e., 12 channels) in the top-ranked algorithms confirms the recommendation of AASM manual in monitoring physiological signals besides EEG. Moreover, the top three ranked entries used LSTM, which has been firmly established in sequence modelling. More detailed information about the challenge and the top-ranked entries can be found in Ghassemi et al (2018).

State-of-the-art CNN architectures have been mainly designed for images and speech signals, but there are few studies with the focus on one-dimensional biosignal classification (e.g., Kiranyaz et al 2016 and Zhang et al 2017). In addition, 1D CNNs are typically faster and more affordable to train compared to recurrent neural networks (RNN), such as LSTM (Bradbury et al 2016). Thus, this study investigates the application of state-of-the-art CNN models to PSG signals. In particular, the main contributions of this paper are:
1) Developing five 1D CNN models that are inspired by the state-of-the-art counterparts mainly designed and used for image or speech processing.
2) Providing empirical studies of the developed models in sleep arousal classification.

The rest of the paper is organized as follows. Section 2 presents the benchmark dataset and preprocessing. Then, 1D CNN models and training procedure are described. The evaluation metric, experimental setup and the evaluation results of the proposed models are presented in Section 3. Section 4 concludes the paper and suggests topics to be explored in further studies.

## 2. Methodology

**2.1 Data preparation**

In this study, we use the dataset provided by PhysioNet CinC Challenge 2018 (PhysioNet 2018). The dataset is collected by the Massachusetts General Hospital's Computational Clinical Neurophysiology Laboratory, and the Clinical Data Animation Laboratory. Multi-channel PSG recordings, including six EEG, one electrooculography (EOG), three electromyography (EMG), one airflow, one oxygen saturation (SaO2), and one electrocardiography (ECG) signals, are recorded with the sampling frequency of 200 Hz. In total, 1893 subjects including 65% male and 35% female with an average age of 55 are monitored. The dataset is partitioned into 994 subjects in the training set and 989 in the testing set. All sleep arousal events were annotated by certified sleep technologists. The labels provided by the organizers contain three distinct values: the target arousals "1", non-arousal regions "0", and regions that are not considered in scoring "-1". The regions are recognized as target arousal if either of the following conditions is met: 1) 2 seconds before beginning and 10 seconds after termination of a respiratory effort related arousal (RERA), and 2) 2 seconds before and after a non-RERA region.

Here, only the five channels of C3-M2 (central EEG), EOG, ABD (EMG of abdominal movement), respiratory airflow, and SaO2 are used. These measurements are selected so that the input of CNNs includes all the different modalities (excluding ECG) during sleep. However, it is worth mentioning that examining various combinations of PSG recordings and analyzing their effects on CNNs' accuracy are beyond the scope of this study.

Each time series is first normalized (between 0 and 1) and, after applying proper anti-aliasing filtering, they are downsampled to 100 Hz. The downsampled signals are windowed into 30 seconds with 25 seconds overlap. Because the original annotations are sample-wise, we reannotate each 30-second window with the most frequent value in the corresponding annotation vector. Afterward, the windows that are marked by "-1" are discarded as they are not being evaluated in the challenge scoring mechanism. Once the classification is performed, the predicted labels are upsampled to have the same size as the original annotations.

**2.2 1D CNN architectures**

In this section, we develop and adapt five different fully 1D CNN models based on the most recent articles. In each architecture, the fundamental elements of designs such as shortcut-connections or dilated convolutions are considered for classification of one-dimensional PSG signals. The architectures and hyperparameters of the 1D CNNs are presented in Fig 1–5 and Table 1.

*Model 1:* The structure of the *first model* is formed by stacking five residual blocks. All the blocks are designed with a shortcut connection to have a more efficient training procedure (He et al 2016). The number of channels needs to be the same in order to add the outputs of shortcut connections with the rest of the blocks. For this purpose, the first block is used as an extra convolution layer (Fig 1. Block A). Each residual block (Fig 1. Block B) consists of two 1D convolutions, and one rectified linear unit (ReLU). The final block is composed of one fully connected layer with hyperbolic tangent (*tanh*) activation function and 150 hidden neurons. In the output layer, the softmax function yields a probability distribution over the class variables. The number of kernels remains the same in all the building blocks and equals to 16. Also, the kernel size in the primary block is 51 and in the residual blocks 25.

*Model 2:* For the *second model* we use FractalNet (Larsson et al 2016). The self-symmetry of the FractalNet is constructed based on the fractal expansion rule:

$$f_{c+1}(z) = [(f_c \circ f_c)(z)] \oplus [conv(z)] \qquad (1)$$

where, $f_1(z) = conv(z)$, ∘ denotes composition, and ⊕ is the merging operation. In this work, dropout is used for regularization instead of drop-path. Similar to *Model 1* the network consists of five Fractalnet blocks (Fig 2). After each block, the feature maps are fed into a max-pooling layer for dimensionality reduction. The final block is composed of two fully connected layers (with 500 and 150 hidden neurons, respectively) with *tanh* activation function. In the output layer, the softmax function is used. The number of kernels in the first block is 8 and increases by 2 in the consecutive blocks (Table 1).

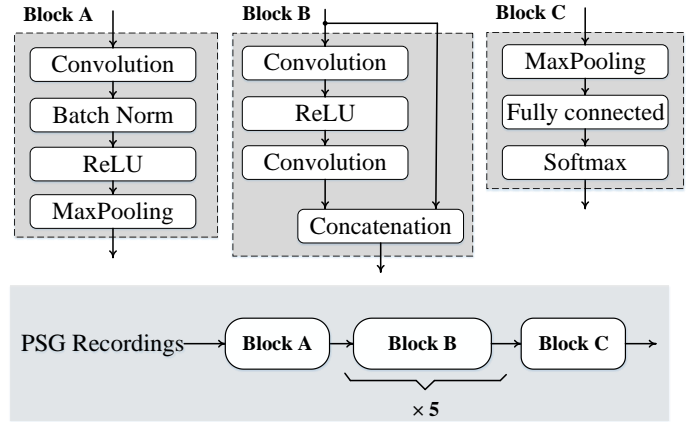

**Figure 1.** The topology of *Model 1*

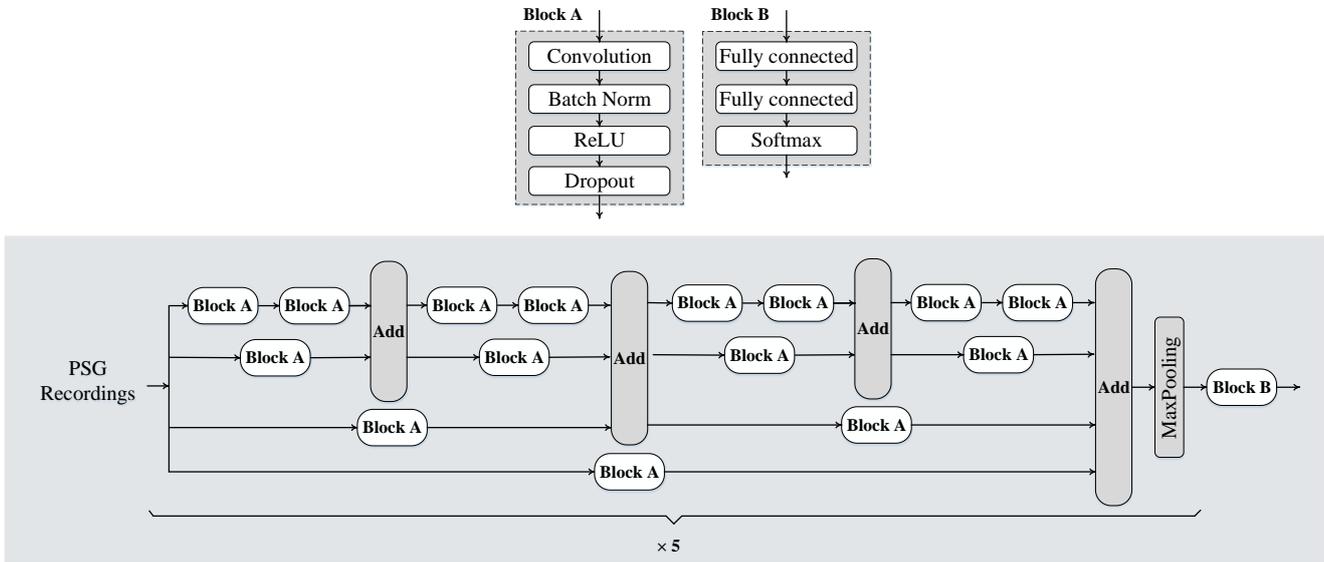

**Figure 2.** The topology of *Model 2*

*Model 3:* The *third model* is chosen based on the 18-layer ResNet (He et al 2016). The network comprises eight consecutive residual blocks with the same topology (Fig 3). In the original ResNet, the number of kernels increases by a factor of 2, but in this work, the number of kernels stays the same (16 and 64) in each block. For the final block, two fully connected layers with tanh activation function, 500 hidden neurons, and a softmax layer are used. The kernel size is similar to the previous architectures, that is 51 in the first block and 25 in the subsequent residual blocks (see Table 1).

*Model 4:* This model is an adaptation of WaveNet (Oord et al 2016) for the binary classification task. WaveNet is designed initially as a deep generative model for speech synthesis and has achieved significant performance compared to the other baseline models in this domain. The two main characteristics of WaveNet are: 1) keeping the autoregressive property (i.e., the output depends on its previous time-steps) using *causal* convolutions and 2) using dilated convolutions to cover a larger receptive field (look-back) without losing resolution. The architecture used in this study is similar to the original one in the article and includes nine dilations (Fig 4. Block B) that stack on top of each other in such a way that the $i^{th}$ layer has a dilation rate of $2^i$. As shown in Fig 4, each block has two outputs, the residual that goes to the next layer, and the skip connections. Skip connections act like shortcut connections to bypass the middle layers and directly connect each layer to the final layer. The Skip connections are merged by adding and passed through a block of ReLU function and average-pooling. The implementation details can be seen in Table 1.

*Model 5:* This model is similar to *Model 4* but with two differences: 1) instead of *causal* convolution we use the ordinary convolution, i.e., with the same padding, and 2) we use only one mutual convolution inside each block (Fig 4. Block B´), which is similar to the original WaveNet paper (Oord et al 2016).

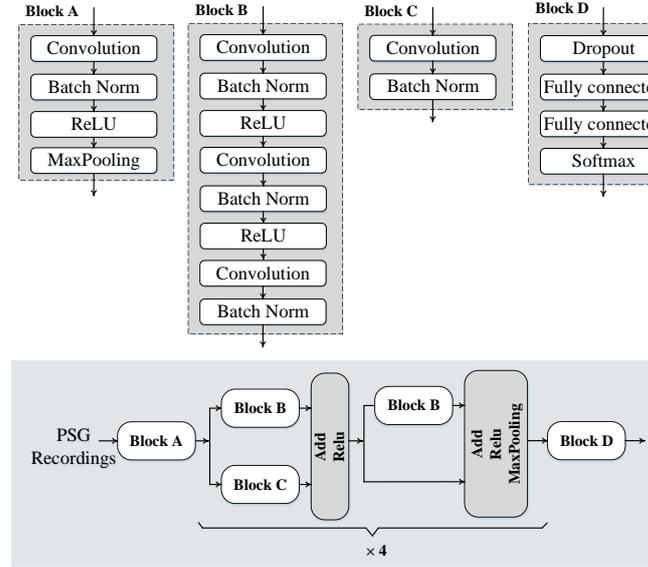

**Figure 3.** The topology of *Model 3*

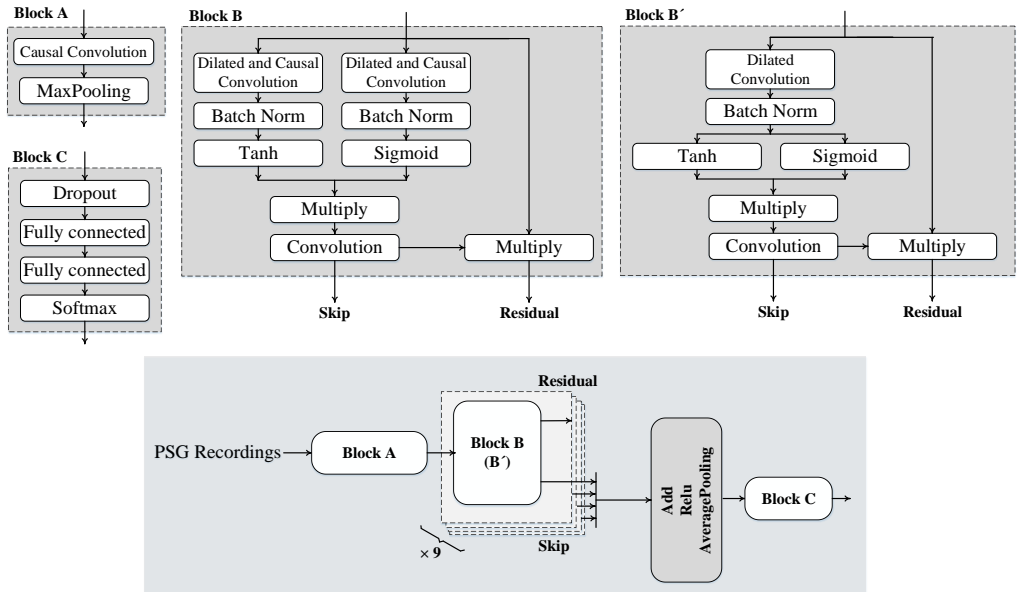

**Figure 4.** The topology of *Model 4* (B) and *5* (B')

### 2.3. Training procedure

The Adam optimization algorithm (P. Kingma and Ba 2014) and cross-entropy loss function are used to train the CNN with 512 mini-batch size, 20 epochs, and a learning rate of 0.001. These hyper-parameters are tuned empirically. To avoid overfitting, we employed an early stopping procedure in such a way that the loss value of the validation set is monitored in each epoch to stop the training procedure if it is not improving within the six previous epochs. For this purpose, 30% of the training set is randomly chosen as the validation set. The model is implemented in Keras (Chollet 2015) using Tensorflow backend (Abadi et al 2015), and the experiments were performed on a workstation with NVIDIA TITAN V GPU and 160 GB memory. Also, due to the high imbalances between arousal (the rare class) and non-arousal events, we balance the data in each mini-batch via stratified sampling, i.e., random sampling of non-arousal class without replacement. The same training strategy and hyper-parameters are used in all the experiments.

**Table 1.** The hyperparameters of the developed architectures.

| Model | Block | Layer | Kernel Size | Number of Kernels | Pool Size | Stride | Padding | Neurons |
|---|---|---|---|---|---|---|---|---|
| 1 | A | Convolution | 1×51 | 16 | - | 1×1 | Valid | - |
| | | Batch Norm | - | - | - | - | - | - |
| | | ReLU | - | - | - | - | - | - |
| | | Max pooling | - | - | 1×2 | - | - | - |
| | B | Convolution | 1×25 | 16 | - | 1×1 | Same | - |
| | | ReLU | - | - | - | - | - | - |
| | | Convolution | 1×25 | 16 | - | 1×1 | Same | - |
| | C | Max pooling | - | - | 1×2 | - | - | - |
| | | Fully connected | - | - | - | - | - | 150 |
| | | Softmax | - | - | - | - | - | 2 |
| 2 | A | Convolution | 51(25) | 8(+2) | - | 1×1 | Same | - |
| | | Batch Norm | - | - | - | - | - | - |
| | | ReLU | - | - | - | - | - | - |
| | | Dropout (0.2) | - | - | - | - | - | - |
| | | Max pooling | - | - | 1×2 | - | - | - |
| | B | Fully connected | - | - | - | - | - | 500 |
| | | Fully connected | - | - | - | - | - | 150 |
| | | Softmax | - | - | - | - | - | 2 |
| 3 | A | Convolution | 1×51 | 16 | - | 1×1 | Same | - |
| | | Batch Norm | - | - | - | - | - | - |
| | | ReLU | - | - | - | - | - | - |
| | | Max pooling | - | - | 1×3 | - | - | - |
| | B | Convolution | 1×1 | 16 | - | 1×1 | Same | - |
| | | Batch Norm | - | - | - | - | - | - |
| | | ReLU | - | - | - | - | - | - |
| | | Convolution | 1×25 | 16 | - | 1×1 | Same | - |
| | | Batch Norm | - | - | - | - | - | - |
| | | ReLU | - | - | - | - | - | - |
| | | Convolution | 1×1 | 16 | - | 1×1 | Same | - |
| | | Batch Norm | - | - | - | - | - | - |
| | | Max pooling | - | - | 1×2 | - | - | - |
| | C | Convolution | 1×1 | 64 | - | 1×1 | Same | - |
| | | Batch Norm | - | - | - | - | - | - |
| | D | Dropout (0.5) | - | - | - | - | - | - |
| | | Fully connected | - | - | - | - | - | 500 |
| | | Fully connected | - | - | - | - | - | 500 |
| | | Softmax | - | - | - | - | - | 2 |
| 4 and 5 | A | Convolution | 1×51 | 16 | - | - | Causal | - |
| | | Max pooling | - | - | 1×2 | - | - | - |
| | B(B´) | Dilated Convolution ($2^i$) | 1×2 | 16 | - | - | Causal (Same) | - |
| | | Batch Norm | - | - | - | - | - | - |
| | | Tanh/Sigmoid/Multiply | - | - | - | - | - | - |
| | | Convolution | 1×1 | 16 | - | - | Same | - |
| | C | Dropout (0.5) | - | - | - | - | - | - |
| | | Fully connected | - | - | - | - | - | 500 |
| | | Fully connected | - | - | - | - | - | 250 |
| | | Softmax | - | - | - | - | - | 2 |

## 3. Results and discussions

To evaluate the performance of the developed architectures, we use the same scoring measure used in the PhysioNet CinC Challenge 2018 (PhysioNet 2018). The area under the precision-recall curve (AUPRC) for the binary classification of arousal and non-arousal regions is defined as follows:

$$AUPRC = \sum_{j,|P_j \cap \bar{N}|} p_j(r_j - r_{j+1}) \qquad (2)$$

where precision is denoted by $p_j = \frac{|A \cap P_j \cap \bar{N}|}{|P_j \cap \bar{N}|}$ and recall is defined as $r_j = \frac{|A \cap P_j \cap \bar{N}|}{|A_j \cap \bar{N}|}$. $N$ and $A$ indicate the set of non-scored samples and target arousal samples, respectively. The reported metric is the gross AUPRC ($AUPRC_G$), which means that the precision and recall are calculated for the entire test/validation database. Moreover, we also report the gross value of the area under the ROC curve ($AUROC_G$).

Table 2 compares the classification results of the developed models in a 3-fold cross-validation scheme. As can be seen in Table 2, the performance of the five models appears to be robust in all the folds. ***Model 1*** obtained the lowest performance among the developed models most likely due to the fact that it is using only one fully connected layer, which reduces the effect of learned features by CNN layers. ***Models 2*** and ***3*** achieve an average AUPRC of 0.29. These two architectures are inspired by FractalNet and ResNet, respectively. The idea behind ***Model 2*** is using sub-paths with different lengths while ***Model 3*** includes residual connections. Based on the empirical results, neither symmetry property, nor residual learning guarantees obtaining a higher performance for arousal detection.

***Models 4*** and ***5*** equally achieve an average $AUPRC_G$ and $AUROC_G$ of 0.31 and 0.84, which yield a superior performance level compared to the competing architectures. These two architectures are the variants of WaveNet for the classification task. WaveNet significantly outperforms prior state-of-the-art techniques in speech synthesis. This confirms that those architectures that perform well in one domain can be successful in other domains as well. As can be seen, ***Model 5*** achieves more robust results than ***Model 4*** (i.e., smaller standard deviations in the cross-validation folds). Thus, for the final model, an ensemble of three models based on ***Model 5*** is designed to be applied to the unseen test data. The three models are trained on different folds and different random seeds. Then, the outputs of the three models are averaged to obtain the final probability.

Furthermore, the cross-validated results reveal that the majority of patients with AUPRC < 1% are the same among all the developed architectures. Obviously, the developed architectures are not capable of learning arousal events in these patients. However, careful observation of the raw signals reveals that there are some confusing class labels. In Fig 5, the signals from C3-M2 channel from two different patients are shown with their designated labels. As can be seen, there is no visible abrupt frequency change in their spectrogram within the target arousal regions. This also applies to the rest of the EEG and EMG channels. In addition, in Fig 6, two small non-arousal interruptions between arousal events from two separate patients are shown. Although label noise is a well-known issue in machine learning problems, these observations raise questions such as: What were the main criteria of annotation? and What was the resolution of the original labels? Providing a detailed description of the annotations will improve the interpretability of results.

There can be two potential reasons for the limited performance of the studied models: 1) using only five channels instead of all, and 2) the constraint on using only fully CNN architecture. In some studies, it has been shown that CNNs can outperform LSTMs in sequential data analysis (e.g., Bai et al 2018). However, the use of LSTM in the classification pipeline of the top three ranked algorithms supports the capability of LSTMs for arousal detection.

**Table 2.** The achieved performance using the five different CNN models.

| Folds | *Model 1* | | *Model 2* | | *Model 3* | | *Model 4* | | *Model 5* | |
|---|---|---|---|---|---|---|---|---|---|---|
| | $AUROC_G$ | $AUPRC_G$ | $AUROC_G$ | $AUPRC_G$ | $AUROC_G$ | $AUPRC_G$ | $AUROC_G$ | $AUPRC_G$ | $AUROC_G$ | $AUPRC_G$ |
| 1 | 0.80 | 0.25 | 0.83 | 0.29 | 0.84 | 0.30 | 0.84 | 0.29 | 0.85 | 0.32 |
| 2 | 0.83 | 0.31 | 0.83 | 0.31 | 0.83 | 0.31 | 0.84 | 0.34 | 0.83 | 0.30 |
| 3 | 0.83 | 0.29 | 0.82 | 0.27 | 0.82 | 0.27 | 0.83 | 0.29 | 0.84 | 0.31 |
| Mean (Standard deviation) | 0.82 (0.02) | 0.28 (0.03) | 0.83 (0.01) | 0.29 (0.02) | 0.83 (0.01) | 0.29 (0.02) | 0.84 (0.01) | 0.31 (0.03) | **0.84 (0.01)** | **0.31 (0.01)** |

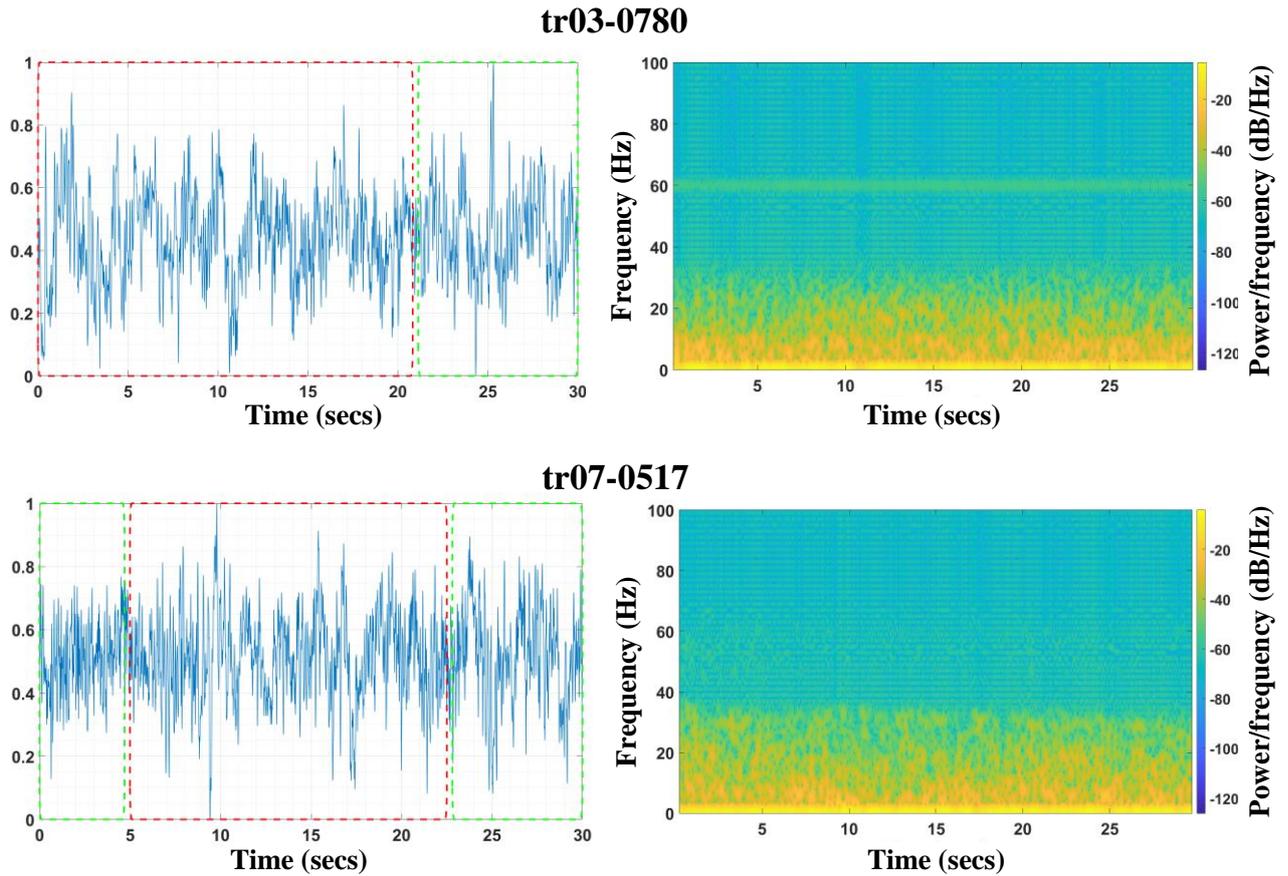

**Figure 5.** The 30-second raw (C3-M2) signals and their spectrograms. The first and second rows belong to patient tr03-0780 and tr07-517, respectively. The target arousal regions, "1", are indicated inside the red rectangle. For the spectrograms, the (Hamming) window length of 100 samples with 85 samples overlap is used.

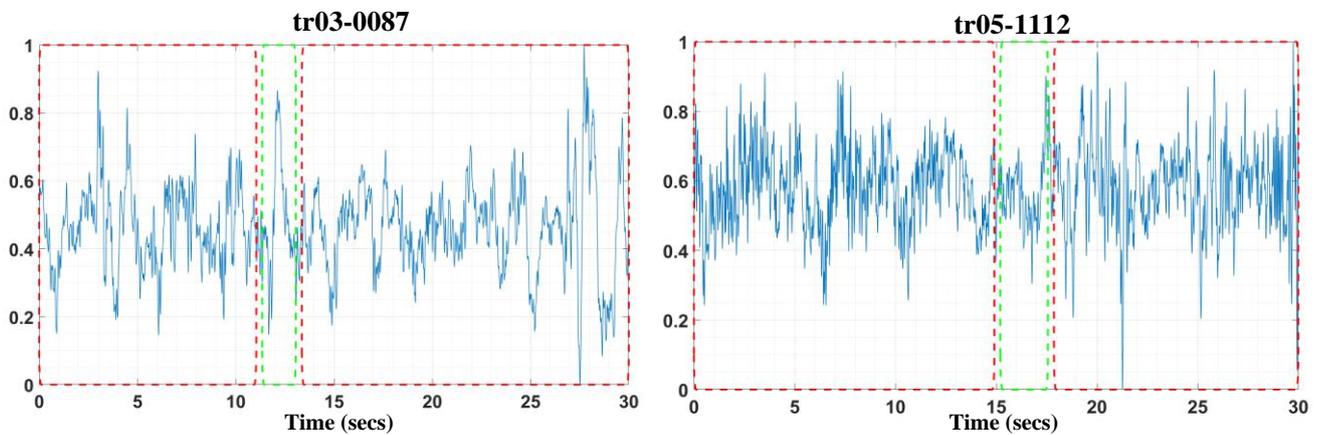

**Figure 6.** Two examples of arousal events with small non-arousal interruptions in C3-M2. The target arousal regions, "1", of patient tr03-0087 (left) and patient tr05-1112 (right) are indicated in the red rectangle.

## 4. Conclusions

Assessment of sleep arousals is a challenging task due to the lack of comprehensive criteria in this domain. The big sleep arousal benchmark dataset provided by the PhysioNet CinC Challenge 2018 has encouraged the use of data-driven algorithms to bridge this gap. Despite the complexity of the problem, the top-ranked algorithms achieved impressive AUPRCs of over 0.4 (maximum 0.54) using deep learning methods. Deep learning models such as CNNs have already been applied in many applications in image and speech processing. However, few studies focus on the adaption of these models for 1D biosignals. In this study, we focus on 1D CNN models and developed five different 1D CNN architectures inspired by the state-of-the-art 2D counterparts. The model performances are then evaluated using the CinC's sleep arousal benchmark dataset.

As the results demonstrate, the adapted models have achieved robust performances in all the folds of cross-validation. Although the trial-error is an essential step for CNN architecture selection, the fundamental elements of designs such as shortcut-connections, dilated convolution, the multiplicity of CNN branches, normalization, and symmetry prove their usefulness for improving the discrimination capability of CNNs in sleep arousals classification. The two variants of WaveNet, *models 4* and *5*, outperform the other architectures with an average AUPRC of 0.31. This shows that the state-of-the-art CNN architectures in one domain can indeed perform successfully in other areas as well. Another observation worth mentioning is that the majority of confusing arousal events (AUPRC <0.1) were mutual among all the developed architectures. The detailed investigation revealed that in these target regions, there is no visible abrupt shift of frequency in the underlying EEG and EMG signals. This makes the interpretations of the underlying reasons for such low AUPRCs difficult. Therefore, providing more detailed information about the dataset can help the interpretability of the results, and consequently, the reduction of false alarms. We expect that experimenting with different window lengths and increasing the number of PSG channels will lead to improvement in the performance of the developed architectures. This will be the topic of our future work.